\documentclass{cflowproc}

\usepackage{natbib}

\begin{document}


\shortauthors{O'SULLIVAN \& VRTILEK}     
\shorttitle{AWM~4 AND MKW~4 OBSERVED WITH XMM} 

\title{AWM~4 and MKW~4 - two very different poor clusters observed with XMM-Newton}   

\author{Ewan O'Sullivan\affilmark{1}   
and J. M. Vrtilek\affilmark{1}}                

\affil{1}{Harvard-Smithsonian Center for Astrophysics}   


\begin{abstract}
We present observations of two poor clusters, AWM 4 and MKW 4, observed by
XMM-Newton. Both systems are relaxed, with little substructure evident in
their X-ray halos or galaxy populations. However, their temperature
structures are markedly different, with AWM 4 isothermal to the resolution
of the EPIC instruments while MKW 4 shows a strong decline in temperature
towards the core. Metal abundance also increases more strongly in the core
of MKW 4 than AWM 4. Three dimensional models show further differences,
suggesting that gas in the core of AWM 4 has been heated and has expanded
outwards. The dominant elliptical galaxy of AWM 4 hosts an AGN with
large-scale radio jets, while the central cD of MKW 4 shows no current AGN
activity. We therefore conclude that the difference in activity cycle of
the AGN in the two galaxies is responsible for the difference in IGM
properties between the two clusters.
\end{abstract}


\section{Introduction}
\label{sec:intro}
MKW~4 and AWM~4 are fairly relaxed poor clusters, originally identified in
the \citet{Morganetal75} and \citet{Albertetal77} surveys. Each is dominated
by a giant elliptical or cD galaxy, surrounded by $\sim$50 (MKW~4) or
$\sim$30 (AWM~4) other galaxies. Kinematic studies of the galaxy
populations of the two clusters show them to be fairly evenly distributed
about their mean redshifts, with no sign of significant substructure
\citep{KoranyiGeller02}. Both clusters are morphologically segregated, with
absorption line systems found predominantly toward the center of each
system. 

The dominant galaxies of MKW~4 and AWM~4 are NGC~4073 and NGC~6051
respectively. Both are highly luminous and $\sim$1.5 magnitudes brighter
than the second ranked galaxy in their clusters. NGC~4073 shows evidence of
a past merger. It has a kinematically decoupled core \citep{Fisheretal95}
and the spectroscopic age of the stellar population is $\sim$7.5 Gyr
\citep{Terlevichforbes00}. It is the larger of the two galaxies, with Log
$L_B$=11.01 $L_\odot$. NGC~6051 hosts an active nucleus, revealed by a
large scale radio feature extending from the galaxy core. The central point
source is not detected at other wavelengths, which suggests that the axis of
the AGN jets is in the plane of the sky and that the central engine is
highly absorbed. The optical luminosity of NGC~6051 is $L_B$=10.76
$L_\odot$.

Both clusters have been the subject of previous X-ray observations with
\textit{Einstein}, \textit{ASCA} and \textit{ROSAT}, all of which have
shown them to be relaxed systems \citep{DellAntonioetal95,JonesForman99}
reasonably well modeled by at most two beta models
\citep{Helsdonponman00,Finoguenovetal01}. The X-ray halo of MKW~4 has a
mean temperature of $\sim$1.7 keV, and radial temperature profiles show
that the temperature increases from $\sim$1.3 keV in the core to a peak
above 2.5 keV at $\sim$115 kpc before dropping back to a fairly constant
temperature of $\sim$1.5 keV at higher radii. The \textit{ASCA} data have
been used to produce radial abundance profiles which show Fe, Si and S
abundances which fall from central values of 0.4-0.7 Z$_\odot$ to 0-0.2
Z$_\odot$ at a radius of 500 kpc \citep{Finoguenovetal00}. The halo of
AWM~4 is somewhat simpler, having a relatively isothermal temperature
profile out to a radius of at least $\sim$200 kpc. The mean temperature of
the gas is $\sim$2.5 keV.

Given their relatively relaxed and undisturbed state, and the large mass
concentrations associated with their central galaxies, these clusters are
potentially excellent sites in which to observe cooling at the low end of
the cluster mass range. We have used two \textit{XMM-Newton} observations
to study the structure of their X-ray halos, and in particular to determine
whether they currently host cooling flows. Throughout these proceedings we
assume $H_0$=75 kms$^{-1}$Mpc$^{-1}$ and normalise optical luminosities to
the B-band luminosity of the sun $L_{B\odot}$=5.2$\times$10$^{32}$ erg
s$^{-1}$. MKW~4 lies at a redshift of $z$=0.020, AWM~4 at $z$=0.032, and
we therefore assume distances to the two clusters of 80 Mpc and 127
Mpc respectively.

\section{Results}
\label{sec:results}

\begin{figure*} 
  \plottwo{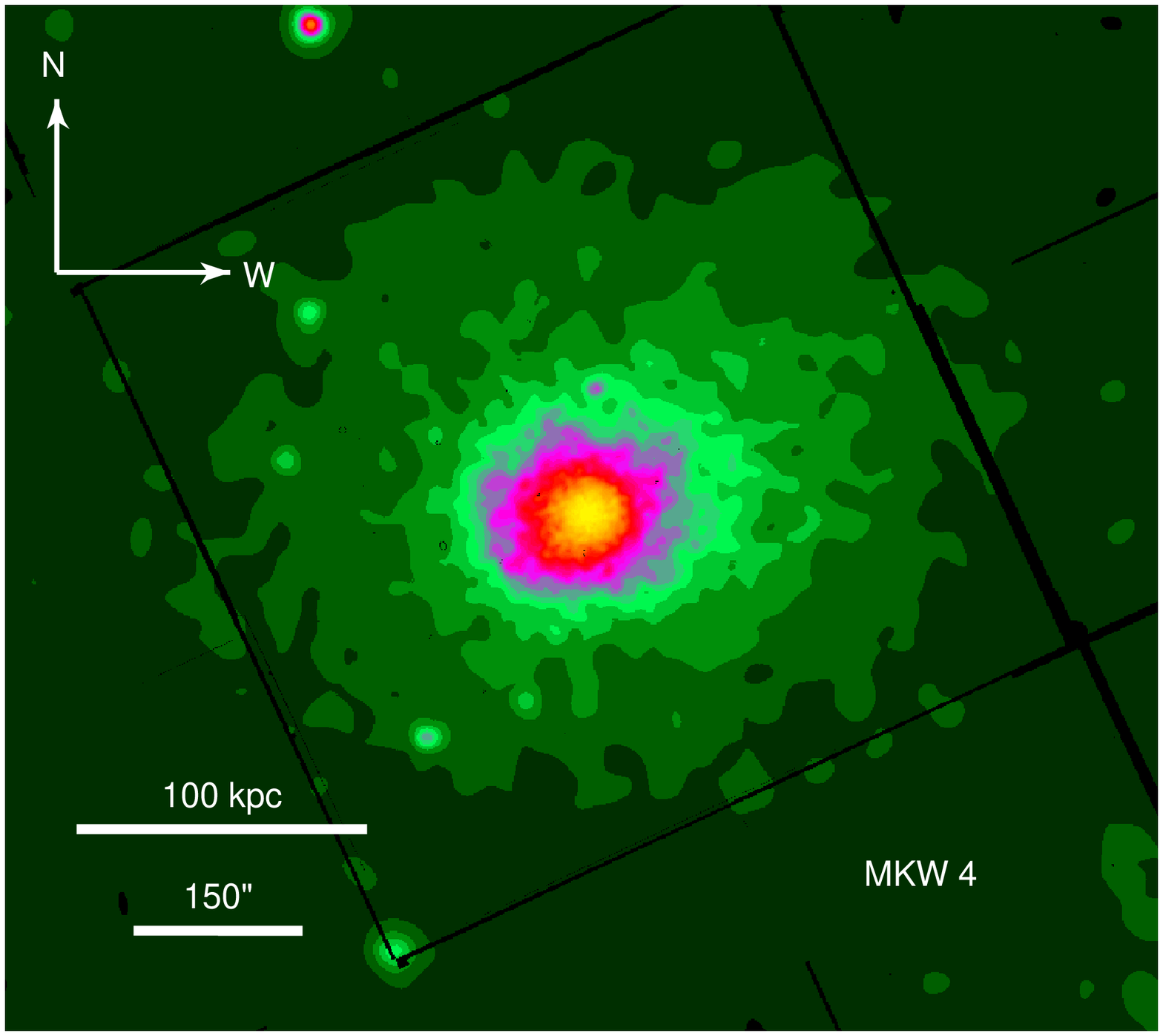}{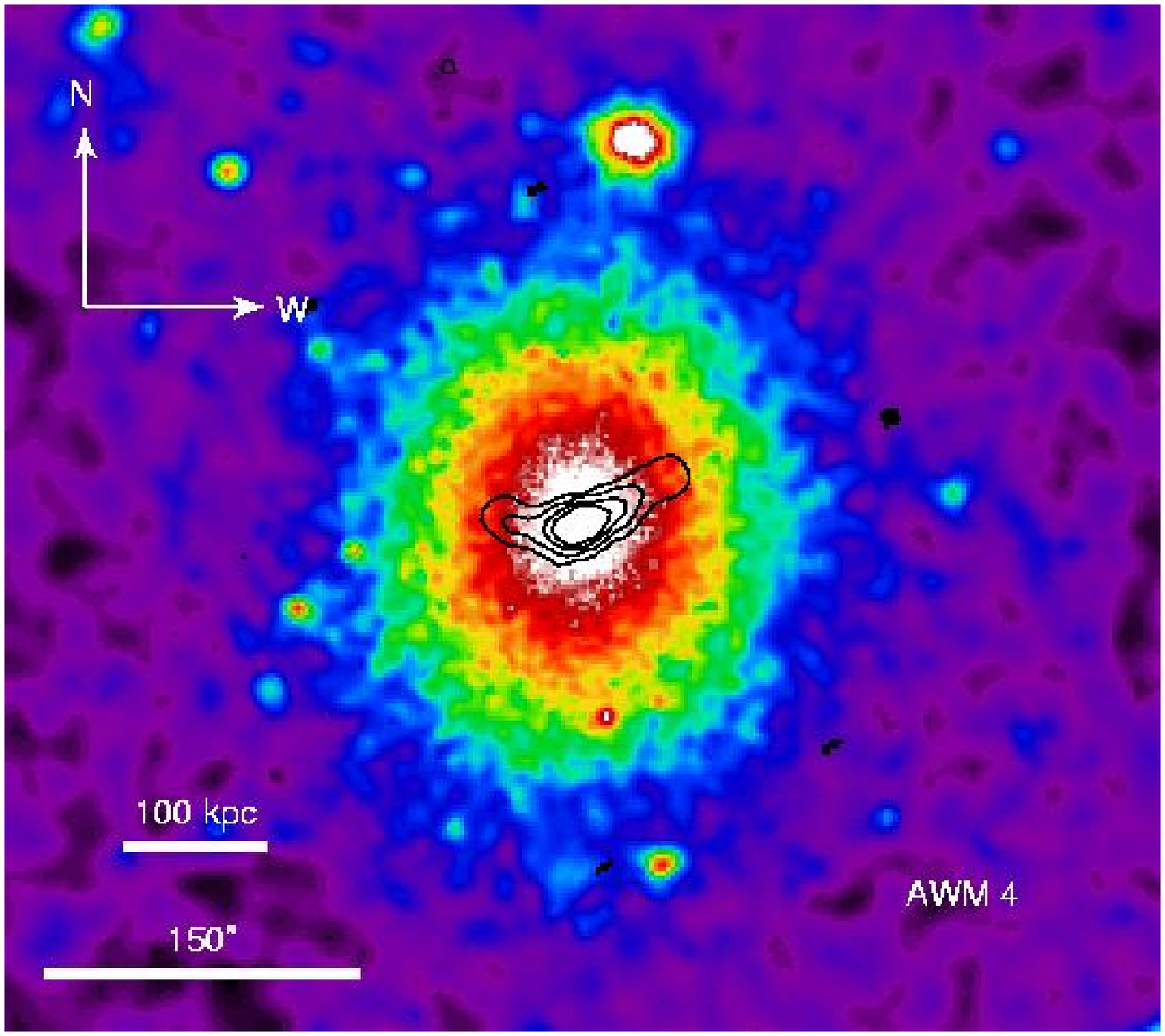}
  \figcaption{\label{fig:images}Adaptively smoothed X-ray images of MKW~4
    (\textit{left}) and AWM~4 (\textit{right}). Radio contours, taken from
    the VLA First 20cm survey, are shown in black on the image of AWM~4.
    The bright source to the north of the AWM~4 cluster is a background
    QSO.}

\end{figure*}

MKW~4 was observed with \textit{XMM-Newton} during orbit 373 (2001 December
21) in two exposures of $\sim$16,000 and $\sim$4,500 seconds. AWM~4 was
observed during orbit 573 (2003 January 25-26) for just over 20,000
seconds. In both cases the EPIC MOS and PN instruments were operated in
full frame and extended full frame modes respectively, with the medium
filter. A detailed summary of the \textit{XMM-Newton} mission and
instrumentation can be found in \citet{Jansenetal01}, and references
therein. The raw data from the EPIC instruments for the longer of the two
exposures were processed with the publicly released version of the
\textit{XMM-Newton} Science Analysis System (\textsc{sas v.5.3.3} for
MKW~4, \textsc{sas v.5.4.1} for AWM~4), using the \textsc{epchain} and
\textsc{emchain} tasks. After filtering for bad pixels and columns, X--ray
events corresponding to patterns 0-12 for the two MOS cameras and patterns
0-4 for the PN camera were accepted.  Investigation of the total count rate
for the field revealed a short background flare at the beginning of the
MKW~4 observation and another in the second half of the AWM~4 observation.
Times when the total count rate deviated above the mean by more than
3$\sigma$ were therefore excluded. The effective exposure times for the MOS
and PN cameras were 14.1 and 10.5 ksec in the case of MKW~4 and 17.4 and
12.7 ksec in the case of AWM~4. Images and spectra were extracted from the
cleaned events lists with the \textsc{sas} task \textsc{evselect}.
Figure~\ref{fig:images} shows adaptively smoothed images of the two
clusters.

\subsection{Surface brightness profiles}
\label{sec:image}
In order to model the surface brightness distributions of the two clusters,
we produced images for use in the \textsc{ciao sherpa} fitting
software. Using data from all three EPIC cameras we searched for and
removed all point sources, with the exception of false identifications
associated with the cluster cores. We also generated background images
based on the blank-sky data of \citet{Lumb02} and \citet{ReadPonman03}, and
the telescope closed data of \citet{Martyetal02}, using the `double
subtraction' technique \citep{Arnaudetal02,Prattetal01}. As the images
contain many pixels with few (or zero) counts, we used the Cash
statistic \citep{Cash79} to perform the fits.

AWM~4 was reasonably well fit by two beta models, a more extended
elliptical component and a circular central component. The parameters (and
1$\sigma$ errors) of the best fit model are shown in Table~\ref{tab:SB}.
Errors for each model were estimated with all fitted parameters free for
perturbation. Position angle is measured anti-clockwise from north.

We initially attempted to model MKW~4 using either a single or
two-component beta model, but this was unsuccessful. We decided instead to
model sectors of the halo individually. We selected four sectors, covering
the major and minor axes of the halo. The NW and SE sectors showed
deviations from a simple beta model, with the NW sector showing emission
extending to higher radii, and the SE sector showing a `bulge' of excess
emission at $\sim$20 kpc. The two remaining sectors were well described by
a single beta model and had similar core radii and slopes, leading us to
fit them together. Taking into account the effect of the PSF and excluding
all data outside the NE and SW sectors, we found the best fit to be a beta
model with parameters as shown in Table~\ref{tab:SB}. Note that this model
is by necessity circular, with the axis ratio fixed at 1.0.

\begin{table*}
\begin{center}
\begin{tabular}{lcccccccc}
Cluster & r$_{core}$ & $\beta_{fit}$ & axis ratio & position angle & amplitude & r$_{core,2}$
& $\beta_{fit,2}$ & amplitude \\
 & (kpc) & & & (degrees) & & (kpc) & & \\
\hline\\[-2mm]
MKW~4 & 4.44$^{+0.15}_{-0.16}$ & 0.447$\pm$0.001 & 1.0 & - & - & - & - & - \\
 & & & & & & & \\[-2mm]
AWM~4 & 96.74$^{+19.13}_{-1.66}$ & 0.708$^{+0.054}_{-0.005}$ &
1.22$^{+0.04}_{-0.02}$ & 174.48$^{+2.39}_{-2.01}$ & 4.155 &
33.58$^{+20.58}_{-9.03}$ & 0.998$^{+0.749}_{-0.347}$ & 7.657 \\
\end{tabular}
\end{center}
\caption{
\label{tab:SB} 
}
\end{table*}    

\subsection{Temperature and Abundance profiles}
\begin{figure*}
\begin{center}
\plotone{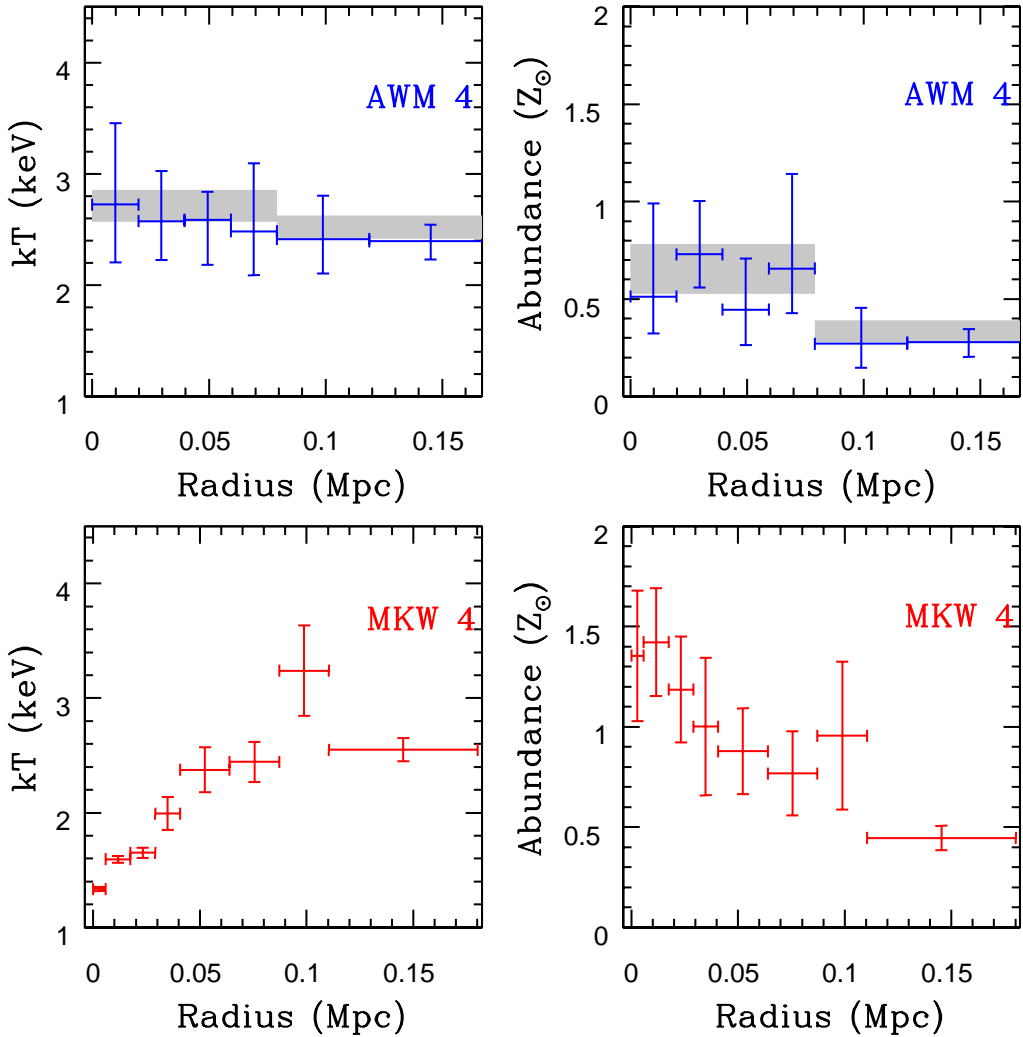}
\end{center}
\caption{\label{fig:Tprof} Deprojected temperature and abundance profiles of the two clusters. Symbols
  represent 90\% error regions. Fitting was carried out using the
  \textsc{xspec} \textsc{projct} model, and assuming the emission arises
  from an absorbed MEKAL model. Hydrogen column was fixed at the galactic
  value. The grey regions in the AWM~4 plots (upper panels) show fits with
  bins tied together to form inner and outer region. Note that the
  temperature difference is still not significant.}
\end{figure*}

In order to examine the large scale temperature and abundance structure of
the clusters we produced radially binned spectra of the halo. Circular bins
were used for MKW~4, elliptical bins with ellipticity and position angle
taken from the surface brightness fits for AWM~4. All data within
17$^{\prime\prime}$ of point sources were removed and spectra were fit
using an absorbed MEKAL model in a 0.4-4 keV energy band.
Figure~\ref{fig:Tprof} shows the results of our fits.

The deprojected temperature profile for MKW~4 shows a fairly smooth decline
in temperature towards the core, consistent with \textit{ROSAT} results
(Helsdon \& Ponman 2000).  Metal abundance is strongly affected by changes
in hydrogen column. The fit quality in the central bins is poor, and we
experimented with different models to fit these bins. The fits could be
improved by using a two-temperature model, but the higher temperature
component was never well constrained by the data. Using cooling flow models
did not produce good fits except in cases where the emission from low
temperature gas (\textit{i.e.} gas at temperatures lower than $\sim$ 1 keV)
was minimal.

The deprojected temperature profile for AWM~4 shows a fairly constant
temperature, with a possible slight decline with increasing radius. To test
whether this decline is real, we tied the four inner and two outer bins to
form two large spectral regions. The results from these are marked as grey
boxes, and show that the decrease is not statistically significant. The
abundance profile shows a more marked decrease with radius. The spectra
from all bins were well described by single temperature plasma models, and
no more complex models were required.

\begin{figure}
\centering
\plotone{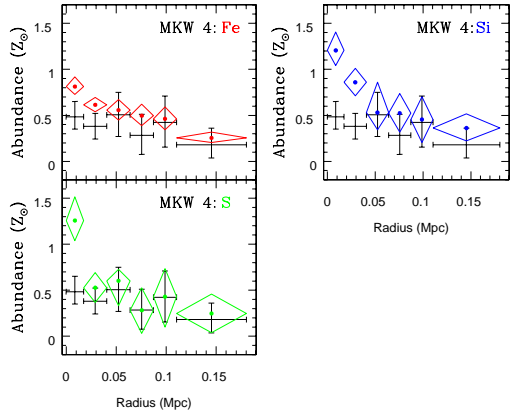}
\caption{\label{fig:MetalsMKW} Radial abundance profiles for Fe, Si and S
  in MKW~4. In each plot, diamonds show the 90\% error on the metal in
  question, as labeled on the plot. The crossed error bars show the fit to
  all the remaining metals, again with 90\% errors. N$_H$ was fixed at the
  galactic value in all fits.}
\end{figure}

\begin{figure}
\centering
\plotone{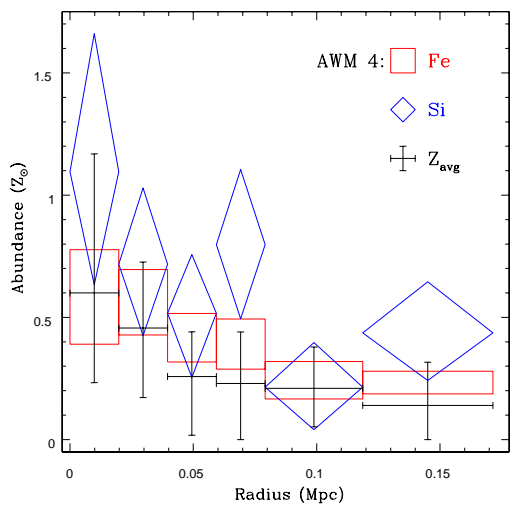}
\caption{\label{fig:MetalsAWM} Radial abundance profiles for Fe and Si in
  AWM~4. Diamonds show the 90\% error on Si, Boxes represent Fe and the
  crossed error bars show the fit to all the remaining metals, again with
  90\% errors. N$_H$ was fixed at the galactic value in all fits.}
\end{figure}

For each cluster, we also investigated the abundances of certain individual
elements, using projected spectra so as to avoid reducing the quality of
the data being fitted.
Using the VMEKAL model in \textsc{xspec} we are able to fit Fe and Si (and
in the case of MKW~4, S)
individually, as well as the abundance of remaining metals and the
temperature in each bin. Figure~\ref{fig:MetalsMKW} shows abundance profiles
for Fe, Si and S in MKW~4, shown with reference to the profile for all other metals
combined. There is a clear increase in all three elements in the core of
the cluster, at the location of NGC~4073. Figure~\ref{fig:MetalsAWM} shows
abundance profiles for Fe and Si in AWM~4, and again shows an increase in
abundance toward the core, at the location of NGC~6051.

Based on the elemental abundances of Fe and Si found in the inner bins of
the two profiles, we are able to estimate the contribution made by SNIa and
SNII to the enrichment of the gas in the cores of the two clusters. We
assume yields of Si and Fe for type II supernovae of
$y_{Fe}$=0.07~M$_\odot$ and $y_{Si}$=0.133~M$_\odot$
\citep{Finoguenovetal00}, and from type Ia supernovae
$y_{Fe}$=0.744~M$_\odot$ and $y_{Si}$=0.158~M$_\odot$
\citep{Thielemannetal93}. For MKW~4 we estimate that $\sim$50\% of the iron
within 17.5 kpc of the core was injected by SNIa. This is similar to our
result for AWM~4, where we find that $\sim$52\% of the iron within 49.2~kpc
was injected by SNIa. However, we note that these values are strongly
affected by our choice of definition of solar abundance ratios. We use the
ratios of \citet{AndersGrevesse79}, but the more recent ratios of
\citet{GrevesseSauval98} give a rather lower value for Fe/H
(Fe/H=3.2$\times$10$^{-5}$ compared to 4.7$\times$10$^{-5}$). This would
suggest that our Fe abundances are underestimated by a factor of $\sim$1.4
compared to those calculated using the more recent ratios, and similarly
that our estimates of SNIa contribution are lower than those we would find
were we to use the Grevesse \& Sauval abundances.

\subsection{Three dimensional gas models}

\begin{figure*}
\plotone{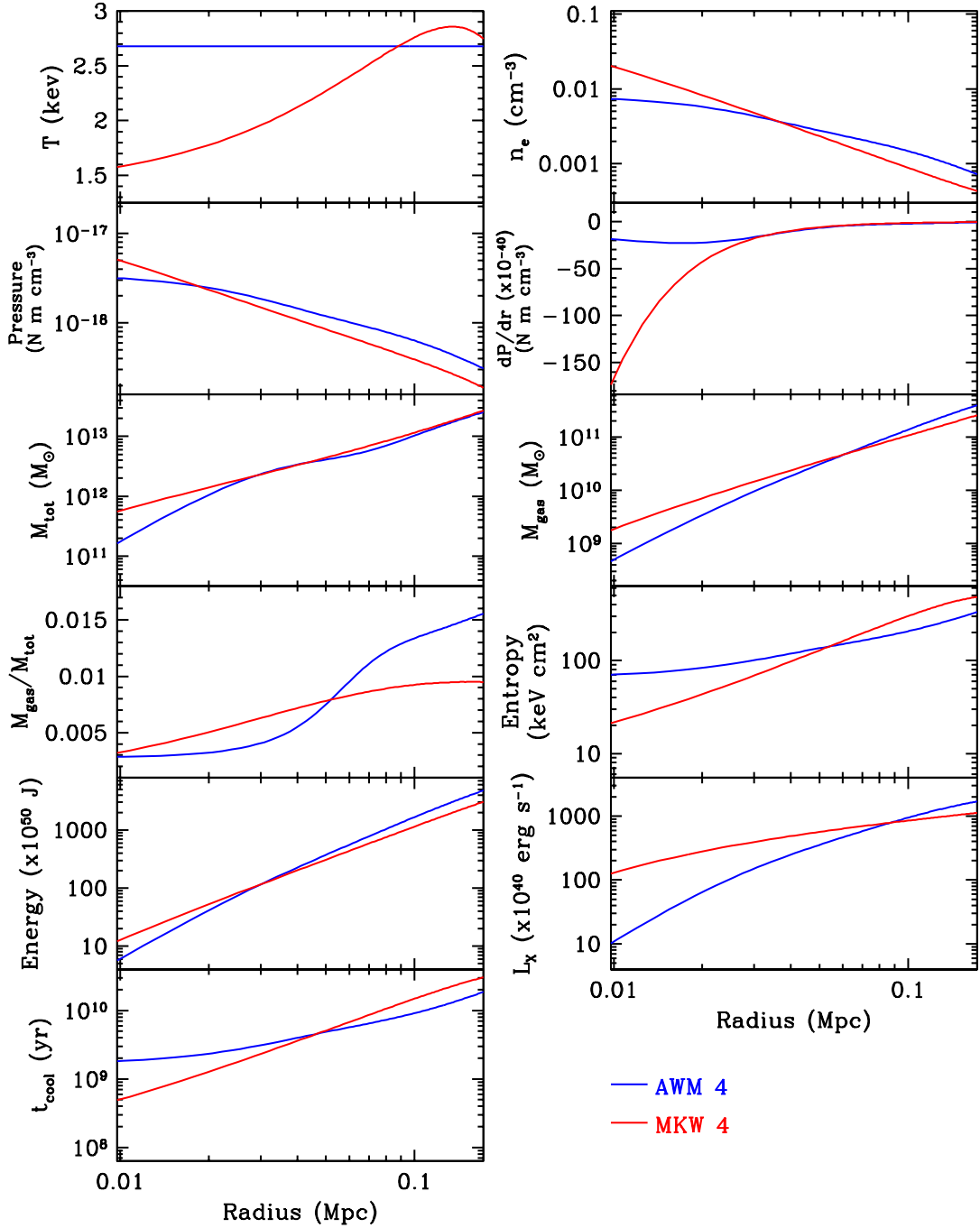}
\caption{Three dimensional properties of the two clusters, as determined
  from their surface brightness and temperature profiles. AWM~4 is shown in
  blue, MKW~4 in red.
\label{fig:entropy}
}
\end{figure*}

Using software provided by S. Helsdon, we use the surface brightness fits
derived in Section~\ref{sec:image} and the radial temperature profiles shown
in Figure~\ref{fig:Tprof} to calculate quantities such as gas mass, total
mass, gas cooling time and gas entropy for the inner region of each cluster.
The software infers the gas density profile based on the fitted surface
brightness profile and a model description of the temperature profile, and
normalises it by comparison with the X--ray luminosity. Given this density
profile we can use the well known equation for hydrostatic equilibrium to calculate the total mass at any given radius. Entropy is defined
as S = T/n$_e^{\frac{2}{3}}$. Figure~\ref{fig:entropy} shows
radial profiles for these and other relevant parameters for both clusters. 

Although the two clusters have quite different structures, their total mass
profiles are very similar outside the core regions. This implies that the
two clusters have similar total masses, with the difference in the core
probably caused by the different masses of the dominant galaxies. Assuming
a mass-to-light ratio for stars of 5~$M_\odot$/$L_\odot$, we would expect
the two galaxies to have total stellar masses of $\sim$5$\times$10$^{11}
M_\odot$ and $\sim$3$\times$10$^{11} M_\odot$, values which are comparable
to the total mass of each system at 10~kpc. MKW~4 has a larger total mass
and a larger mass of gas in its core regions than AWM~4. The gas fraction
of AWM~4 is very low out to $\sim$40 kpc, where it begins to increase
rapidly. In the outer regions the situation is reversed, and AWM~4 has a
considerably larger gas fraction than MKW~4.  MKW~4 has a relatively short
cooling time in the core ($\sim$5$\times$10$^{8}$ yr), and the inner halo
contains very low entropy gas. In AWM~4, gas entropy is significantly
higher, and the cooling time is relatively long, even at very small radii.
This suggests that while MKW~4 has a cool core, and may be developing
towards a cooling flow, the core of AWM~4 has likely been heated. The gas
fraction profile suggests that the inner halo has been `puffed up' by the
injection of energy, which is also responsible for the isothermal
temperature profile.

\section{Discussion and Conclusions}
MKW~4 is similar to a number of systems observed by \textit{Chandra} and
\textit{XMM} in that although its halo contains gas at a range of
temperatures, there is no evidence of cooling below $\sim$0.5 keV. NGC~4636
\citep{Xuetal02}, M87 \citep{Sakelliouetal02} and NGC~5044
\citep{Tamuraetal03} all show this behaviour, with minimum temperatures of
0.5-0.6 keV. Given this minimum temperature in MKW~4, it seems likely that
the halo has been heated in the past and is now in the process of cooling
and re-establishing a cooling flow. This heating could have been caused by
some dynamical event, such as a sub-cluster merger, or by AGN activity.

We calculated the time required for an isothermal halo to cool to the point
where its temperature profile would match that which we observe in MKW~4.
In the cluster core, this cooling timescale is $\sim$200 Myr. This is quite
comparable to the activity cycle of an AGN, but very dissimilar to the
spectroscopic age of the dominant galaxy. Along with the relaxed nature of
the cluster, this rules out a subcluster merger as the source of heating.
We therefore suggest that MKW~4 is likely to have been heated by an AGN
which is currently quiescent. We also note that the temperature profile of
MKW~4 is not well described by either the `universal' temperature profile
for relaxed clusters of \citet{Allenetal01} or by models in which cooling
is balanced by conduction \citep{Voigtetal02}. This may support our
suggestion that MKW~4 is not in a stable state and is likely to cool
further.

AGN heating is also likely responsible for the isothermality of AWM~4.
Taking the unlikely possibility that the isothermal temperature profile we
observe is a stable, long term feature of the system, an energy input of at
least $\sim$10$^{43}$ erg s$^{-1}$ is required to balance the X-ray
emission from the central 100 kpc. If, as is more likely, AWM~4 once had a
temperature profile much like that of MKW~4, $\sim$9$\times$10$^{58}$ erg
would be needed to heat it to its current state. This is equivalent to an
AGN injecting $\sim$3$\times$10$^{43}$ erg s$^{-1}$ for 100 Myr. We
therefore believe that even considering efficiency factors, it is
realistic to assume AWM~4 has been heated by AGN activity. 

In conclusion, we find that AWM~4 and MKW~4 are very similar systems with
two major differences:

\begin{enumerate}
\item \textbf{AGN activity:} While AWM~4 is currently being
heated by the AGN in NGC~6051, there is no ongoing AGN activity in
NGC~4073, and gas in the core of MKW~4 has been able to cool.
\item \textbf{Galaxy mass:} NGC~6051 is considerably less massive
than NGC~4073, leading to a less steeply peaked mass profile in the core of
AWM~4.
\end{enumerate}

The observed differences in temperature and surface brightness profile
between the two systems can be explained largely through the effects of
their respective central galaxies. The mass of these galaxies and the
current phase of their AGN duty cycles appear to have profound effects on
the structure and future evolution of the two clusters.


\acknowledgements
We are very grateful to S. Helsdon for the use of his software. This work was supported in part by NASA grants NAG5-10071 and GO2-3186X.

\bibliography{../paper.bib}
\bibliographystyle{apj}



\end{document}